\documentclass[pre,aps,amssymb,twocolumn,groupedaddress,notitlepage,superscriptaddress,longbibliography]{revtex4-1}
\RequirePackage[mathlines]{lineno}
\usepackage{amssymb}
\usepackage{amsmath}
\usepackage{graphicx}
\usepackage[usenames,dvipsnames]{xcolor}
\definecolor{goodgreen}{rgb}{0.1,0.5,0}
\definecolor{goodred}{rgb}{0.7,0,0}
\usepackage[colorlinks,urlcolor=goodgreen,citecolor=blue,linkcolor=goodred]{hyperref}

\makeatletter
\newsavebox{\@brx}
\newcommand{\llangle}[1][]{\savebox{\@brx}{\(\m@th{#1\langle}\)}%
  \mathopen{\copy\@brx\kern-0.5\wd\@brx\usebox{\@brx}}}
\newcommand{\rrangle}[1][]{\savebox{\@brx}{\(\m@th{#1\rangle}\)}%
  \mathclose{\copy\@brx\kern-0.5\wd\@brx\usebox{\@brx}}}
\makeatother

\usepackage[normalem]{ulem}

\begin{document}

\title{Optical Hall response of bilayer graphene: the manifestation of chiral hybridized states in broken mirror symmetry lattices}

\author{V. Nam Do}
\email{nam.dovan@phenikaa-uni.edu.vn}    
\affiliation{Phenikaa Institute for Advanced Study (PIAS), A1 Building, Phenikaa University, Hanoi 10000, Vietnam}
\author{H. Anh Le}
\affiliation{Phenikaa Institute for Advanced Study (PIAS), A1 Building, Phenikaa University, Hanoi 10000, Vietnam}
\author{V. Duy Nguyen}
\affiliation{Phenikaa Institute for Advanced Study (PIAS), A1 Building, Phenikaa University, Hanoi 10000, Vietnam}
\author{D. Bercioux}
\email{dario.bercioux@dipc.org}
\affiliation{Donostia International Physics Center (DIPC), Manuel de Lardizbal 4, E-20018 San Sebast\'ian, Spain}
\affiliation{IKERBASQUE, Basque Foundation of Science, 48011 Bilbao, Basque Country, Spain}
\begin{abstract}
Understanding the mechanisms governing the optical activity of layered-stacked materials is crucial to the design of devices aimed at manipulating light at the nanoscale. Here, we show that both twisted and slid bilayer graphene are chiral systems that can deflect the polarization of linear polarized light. However, only twisted bilayer graphene supports circular dichroism. Our calculation scheme, which is based on the time-dependent Schr\"odinger equation, is particularly efficient for calculating the optical-conductivity tensor. Specifically, it allows us to show the chirality of hybridized states as the handedness-dependent bending of the trajectory of kicked Gaussian wave packets in bilayer lattices. We show that nonzero Hall conductivity is the result of the noncanceling manifestation of hybridized states in chiral lattices. We also demonstrate the continuous dependence of the conductivity tensor on the twist angle and the sliding vector.
\end{abstract}

\maketitle
\section{Introduction}

Stacked two-dimensional (2D) materials represent a unique platform for the  manipulation of light at the nanometer scale, therefore, they are also an  ideal platform for advances in future emerging technologies~\cite{Wang-2010, Havener-2014,Gupta-2018,Akinwande_2019}.
Additionally, twisted stacks of graphene or other so-called 2D van-der-Waals materials  may be used torealize twistronics and optoelectronics devices based on tuning their electronic  structure~\cite{Carr-2017,Cao-2018a,Cao-2018b,Palau-2018,Seyler-2019,Yankowitz-2019,Saito_2020,Hu_2020}.
It was predicted that using  Bernal-stacked bilayer graphene  may produce a finite Faraday rotation  of light when traveling through a microcavities~\cite{Da-2016}. In recent experiments, it was demonstrated that twisted bilayer graphene (TBG) can be used to manipulate the polarization state of light, resulting in a finite circular dichroism (CD)~\cite{Kim-2016}. Since intrinsic monolayer graphene does not have this property, understanding how twisting layers of graphene results in a finite optical activity, besides being a fundamental physical question is also essential for the development of nanodevices with novel chiral properties~\cite{Kim-2016} with  important applications for recognizing different enantiomers of molecules~\cite{Izake_2007,Liu_2005}.

Usually, the application of a magnetic field leads to the generation of finite Faraday and Kerr rotations of the light polarization plane~\cite{Tse-2011,Crassee-2011,Nandkishore-2011,Shimano-2013}. However, this requires large devices, thus their practical applications are limited. It was found that an electronic ground state with broken time-reversal symmetry (TRS) may support the appearance of Faraday rotation in the absence of a magnetic field~\cite{Gorba-2012,Szechenyi-2016}.  A strained graphene lattice was also shown to exhibit giant Faraday and Kerr rotation~\cite{Martinez-2012,Leyva-2015}.  The electronic structure of strained graphene can be described by a picture of Weyl-like fermions moving in a gauge field~\cite{Levy-2010,Juan-2013}.  It was also pointed out that the physics of low-energy electronic states in TBG is governed by an effective non-Abelian gauge field~\cite{Jose-2012,Ramires-2019}.  Similarly to spin-orbit interactions~\cite{Bercioux_2015}, these non-Abelian gauge fields preserve TRS~\cite{Jose-2012}. An alternative analysis is terms of a Berry curvature dipole was recently proposed~\cite{Battilomo_2019}. It has been predicted that the deformation of electron states caused by twisting and sliding graphene layers will manifest  through unique transport and optical properties such as a nonzero optical Hall response and the anisotropy of  longitudinal conductance~\cite{Guinea-2010,son-2011,Pellegrino-2010,Leyva-2016,Ochoa_2020}.

Phenomenological models are usually employed to describe the optical activity of solids; see Refs.~\cite{Stauber-2018, Dung-2020,Ochoa_2020} for recent proposals describing the low-frequency regime of the chiral response in chiral 2D materials. However, a microscopic approach was recently proposed by Su\'arez Morell~\emph{el~al.} in Ref.~\cite{Morell-2017}. Here, the analysis of the optical activity is based on the decomposition of the current operator into  components in each graphene layer. They introduced an external parameter describing the phase factor characterizing the dephasing of the currents in two different graphene layers. The optical Hall conductivity is then deduced as a result of the correlation between the current components in the two layers. They concluded that the relative rotation of the electron chirality due to the lattice twisting and the current dephasing are the origin of the circular dichroism of the TBG system. 

In this work, we present a microscopic analysis for the optical activity of TBG.
We find that  decomposing the current into  different contributions from the two layers is not a conclusive interpretation for describing the optical activity of TBGs and the other bilayer graphene (BLG) systems~\cite{Morell-2017}. Instead, we notice a crucial role played by the electron dynamics in the twisted or slid lattices. Under twisting or sliding, the change of the lattice symmetry induces the spatial deformation of the wave functions of hybridized electron states. It is thus responsible for the chiral response of the bilayer graphene systems. 
We introduce an efficient  scheme to calculate all elements of the optical conductivity tensor rather than only  the longitudinal conductivities. Our approach allows us to conside the electron dynamics at the atomic scale with respect to all the natural symmetries of the atomic lattice~\cite{Zou-2018}. 
Secondly, we show how the optical Hall response of the system is governed not only by the inter-layer current-current correlations as pointed out by Su\'arez Morell~\emph{et al.}, but  by the intra-layer current-current correlations as well.

In the following, we show that only hybridized states formed by electrons between the two layers govern the optical activity of the bilayer system. These hybridized states support the electron propagation not only in each graphene layer but also interchangeably between the two layers~\cite{Xian-2013,Do-2019}. However, their contribution to the Hall response depends on their spatial symmetries. We show that when the mirror symmetry is broken, the hybridized states have no cancelling contribution to the optical Hall conductivity, resulting in a nonzero value for this quantity. 
Our analyses are based on a real-space approach, entirely at the microscopical level. To study the optical Hall response, we express the Kubo formula for the conductivity tensor in the form of the Kubo-Bastin formula. On a practical level, we obtain the conductivity tensor within the Kernel Polynomial Method (KPM)~\cite{Weibe-2006,Le-2018,Le_2019,Do-2019}.  This numerical approach allows us to work with arbitrary configurations of the bilayer graphene, \emph{i.e.}, taking into account both the twist angle and the sliding vector, and considering all of the natural symmetries of the bilayer atomic lattice. We do not need to find the electronic eigenfunctions explicitly: we performed the calculation based on the analysis of the time-evolution of two kinds of states in the bilayer lattice | the localized $2p_z$-states and the kicked Gaussian wave packets. We show that the trajectory of the centroid of wave packets deviate from the direction of the initial wave vector and, the deviated direction depends on the initial layer location of the wave packet. This demonstrates the transverse correlation of the electron motion and hence, the dependence of the Hall response on the chirality of the bilayer lattice.

The paper is organized as follows: In Sec.~\ref{secII}, we present a model used to describe the dynamics of electrons in the BLG lattice together with the Kubo-Bastin formula. In Sec.~\ref{secIII}, we discuss the optical Hall response of the BLG configurations through the analysis of the behavior of the optical conductivity components. In Sec.~\ref{wpd}, we present results illustrating the wave-packet dynamics in single layer and BLG systems. We devote Sec.~\ref{secIV} to a discussion of the optical activity  of the BLG system through a determination of the Faraday and Kerr rotation angles as well as the circular dichroism.  Finally, our conclusions are given Sec.~\ref{secV}. A few technical appendices  comple the manuscript. In~\ref{appA} we give details on the Kubo-Bastin formula for the conductivity tensor and its evaluation in terms of the KPM. In~\ref{appB}, we provide a highlight of the representation of relevant operators in terms of Chebyshev polynomials. In~\ref{appC} we highlight the relation between the components of the conductivity tensor and the optical coefficients.

\section{Model and method}\label{secII}

To characterize the dynamics of electrons in the BLG system we use  
a microscopic approach based on a tight-binding Hamiltonian describing electrons in the $2p_z$ orbitals of carbon atoms. The system Hamiltonian reads~\cite{Moon-2013,Koshino-2015,Le-2018,Do-2019,Le_2019}:
%
%
\begin{equation}\label{Eq5}
\hat{\mathcal{H}}=\sum_{\nu=1}^2\sum_{i,j}t_{ij}^\nu |\nu i\rangle\langle\nu j|+\sum_{\nu\neq\bar{\nu}=1}^2\sum_{ij}t_{ij}^{\nu\bar{\nu}}|\nu i\rangle\langle\bar{\nu} j|.
\end{equation}
%
%
Here, the first term defines the dynamics of an electron in each of the monolayer labeled by the index $\nu$ from site $i$ to site $j$ with the intra-layer hopping energy $t_{ij}^{\nu}$; the basis set is given by the ket-states $\{|\nu i\rangle\}$ representing the $2p_z$-orbitals of  carbon atoms. The second term in Eq.~(\ref{Eq5}) describes the electron hopping between two layers which is characterized by the hopping parameters $t_{ij}^{\nu\bar{\nu}}$. We use the Slater-Koster formalism to determine the values of the hopping parameters $t^\nu_{ij}$ and $t^{\nu\bar{\nu}}_{ij}$~\cite{Moon-2013, Koshino-2015,Do-2019}. In this work, we will ignore the effects of the graphene sheet curvature~\cite{Uchida-2014,Lucignano-2019,Cantele_2020}; we assume the spacing between the two layers constant and about $d\approx 3.35$~\AA\, and we set all onsite energies to be zero. We will treat the BLG system in the general form by considering two different types of configurations: (a) twisted bilayer graphene, and (b) slid bilayer graphene (SBG).
In the case (a), the two layers are rotated with respect to each other by a twist angle $\theta$. In general, in this configuration the system does not have translational symmetry, but supports moir\'e patterns | a typical feature of TBG configurations. The translational symmetry is only recovered for a discrete, but infinite set of twist angles given by the expression: 
%
%
\begin{equation*}
\cos\theta = \frac{3q^2-p^2}{3q^2+p^2},
\end{equation*}
%
%
where $p$ and $q$ are integers~\cite{Shallcross-2008}. When the twist angle $\theta$ satisfies this equation, the staking of two monolayer lattices is called commensurate, otherwise it is incommensurate. The unit cell of commensurate TBG configurations with tiny twist angles usually contains thousands of carbon atoms, causing limitations in the calculation using  exact diagonalization procedures.  In contrast, in configuration (b),  translational symmetry is preserved, but the point group symmetries are changed compared to the case without sliding. The unit cell is always defined in this configuration and it is composed of four carbon atoms, two from each layer~\cite{son-2011,Jose-2012}. 

The key to  theoretically studying the Hall response of an electronic system is to calculate and analyze  the electrical conductivity tensor. In linear response theory, there are several formulations for the Kubo conductivity suitable for calculating either the longitudinal conductivities or the transversal ones~\cite{Roche-1997,Yuan-2010,Ortmann-2015,Ferreira-2015}. Starting from a real-space approach, we aim to calculate all the elements of the conductivity tensor within a unique formalism that also works for systems lacking translational  invariance~\cite{Wang-1994,Garcia-2015, Le-2018}. Specifically, we use the following expression to calculate the conductivity tensor, also known as the Kubo-Bastin formula~\cite{Bastin-1971,Garcia-2015}:
%
%
\begin{align}\label{Eq1}
\sigma_{\alpha\beta}(\omega) &= \frac{ie^2}{\omega}\frac{1}{\Omega}\int_{-\infty}^{+\infty}dEf(E)\nonumber\\
&\hspace{1cm}\times\text{Tr}\left\{\delta(E-\hat{\mathcal{H}})\hat{v}_\alpha\hat{\mathcal{G}}^+(E+\hbar\omega)\hat{v}_\beta\right.\nonumber\\
&\hspace{1.6cm}\left.+\hat{\mathcal{G}}^-(E-\hbar\omega)\hat{v}_\alpha\delta(E-\hat{\mathcal{H}})\hat{v}_\beta\right\},
\end{align}
%
%
where $\hat{\mathcal{G}}^{\pm}(E) = (E-\hat{\mathcal{H}}\pm i\delta)^{-1} $ are the retarded ($+$) and advanced ($-$) resolvents, respectively, and $\hat{v}_{\alpha}= i\left[\hat{\mathcal{H}},\hat{x}_\alpha\right]/\hbar$ is the $\alpha$-component of the velocity operator. In~\ref{appA}, we present the derivation for Eq.~(\ref{Eq1}) starting from the more general Kubo formula. We implement Eq.~(\ref{Eq1}) within the KPM by using the Chebyshev polynomials of the first kind, $T_m(x) = \cos[m\mathrm{acos}(x)]$, to represent the operators:
%
%
\begin{subequations}\label{neweq_3}
\begin{align}
&\delta(E-\mathcal{\hat{H}}) = \frac{\theta(1-\epsilon)\theta(1+\epsilon)}{W\pi\sqrt{1-\epsilon^2}}\sum_{m=0}^{\infty}\frac{2}{\delta_{m,0}+1}T_m(\epsilon)T_m(\hat{h}),\label{Eq2}\\
&\hat{\mathcal{G}}^{\pm}(E) = \frac{1}{W}\sum_{m=0}^\infty \frac{2}{\delta_{m,0}+1}(\mp i)^{m+1}g_m^\pm\left(\epsilon\pm i\eta\right)T_m(\hat{h})\label{Eq3}.
\end{align}
\end{subequations}
%
%
In the previous expressions, we have rescaled the energy variable and the Hamiltonian in the range of $(-1,1)$: 
%
%
\begin{eqnarray*}
E\to\epsilon  &=& \frac{E-E_0}{W}\\
\hat{\mathcal{H}}\to\hat{h} & = &\frac{\hat{\mathcal{H}}-E_0}{W},
\end{eqnarray*}
%
%
where $W$ is the half of spectrum bandwidth, $E_0$ is the central point of the spectrum. The function $g_m(z)$ is defined by
%
%
\begin{equation}
g_m^\pm(z) = \frac{1}{\sqrt{1-z^2}}\left(\sqrt{1-z^2}\pm iz\right)^m\label{Eq4}
\end{equation}
%
%
with complex variable $z$ taking the values as $z_\pm=\epsilon\pm i\eta$ to define the resolvents $\hat{\mathcal{G}}^\pm$. 

Substituting expressions~(\ref{neweq_3})  into Eq.~(\ref{Eq1}) leads a calculation of the so-called Chebyshev momenta
%
%
\begin{equation}
\chi_{mn} = \mathrm{Tr}[T_m(\hat{h})\hat{v}_\alpha T_n(\hat{h})\hat{v}_\beta].
\end{equation}
%
%
These quantities are commonly evaluated by stochastic methods with the use of a set of random phase states~\cite{Weibe-2006}. In our work, we use the scheme of randomly sampling the basis set to build a small set of $|\nu i\rangle$~\cite{Le-2018,Do-2019}. When adopting this set of states, the Chebyshev momenta $\chi_{mn}$ are simply evaluated by $\chi_{mn}=\sum_{i}\chi_{mn}^{(i)}$ where $\chi_{mn}^{(i)} = \langle\nu i|T_m(\hat{h})\hat{v}_\alpha T_n(\hat{h})\hat{v}_\beta|\nu i\rangle$.  
One of the advantage of this technique is that  it avoids special treatments of nodes near the sample edges, which are usually affected by boundary conditions imposed by calculation. Additionally, it allows to interpret the final result as the contribution of local information on each lattice site in particular domains of the lattice, {\it e.g.}, the unit cell or the moir\'e cell in the TBG system. 
%
%
\begin{figure*}[!t]\centering
\includegraphics[width = \textwidth]{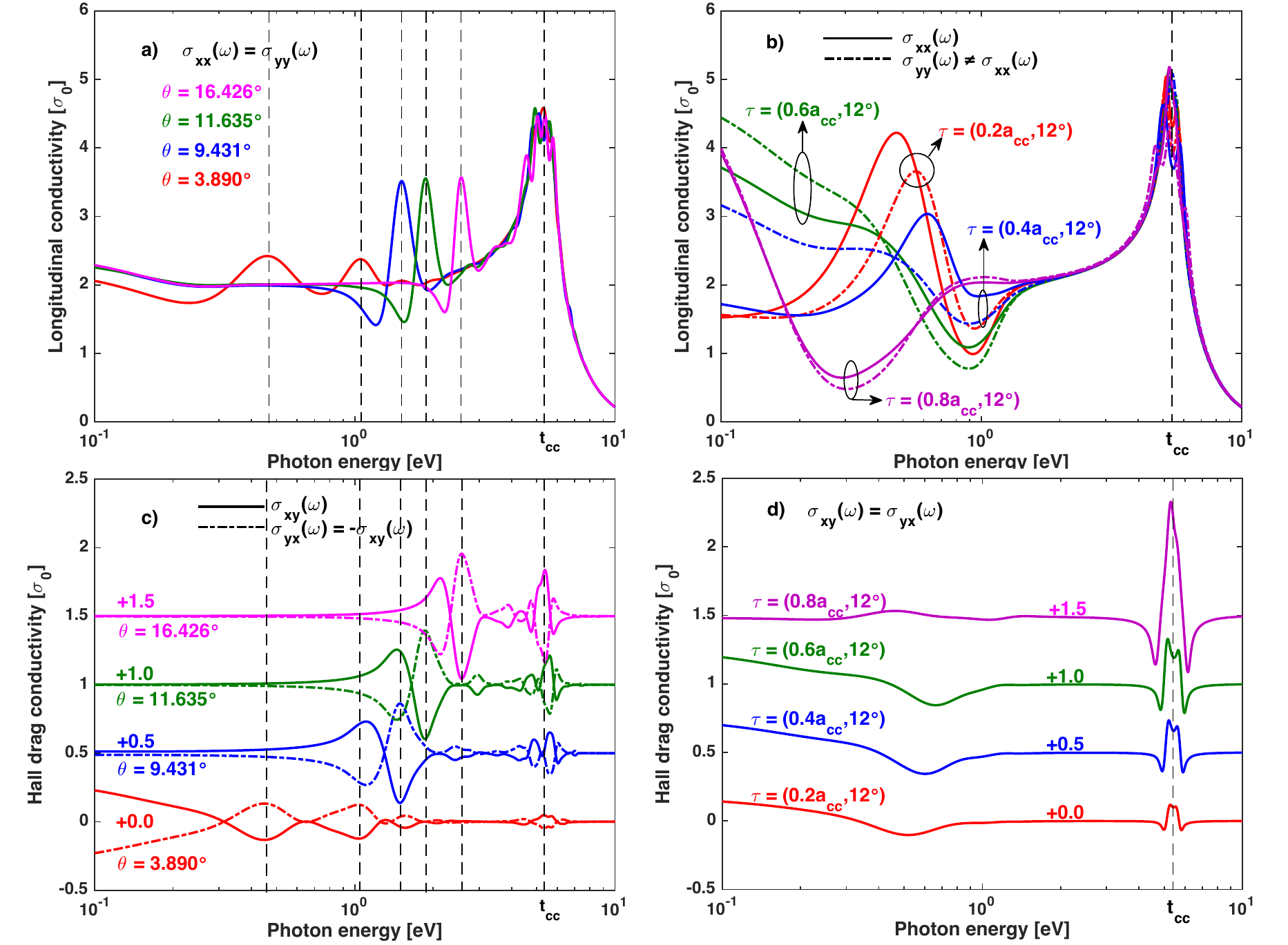}
\caption{\label{fig_1}Real part of the longitudinal $\sigma_{xx}(\omega)$ and transversal optical Hall conductivity $\sigma_{xy}(\omega)$ as a function of the photon frequency
for several TBG [a) \& c)] and SBG [b) \& d)] configurations. The conductivities are expressed in the unit of $\sigma_0 = e^2/4\hbar$. The conductivity presented in the panels c) and d) are shifted upward to distinguish the curves. In the four panels, the vertical dashed lines are added to highlight the position of the conductivity peaks.}
\end{figure*}
%
%

\section{Optical Hall response}\label{secIII}

We present in Fig.~\ref{fig_1}a) results of the optical conductivity tensors for four TBG configurations with the following twist angles $\theta = 16.426^\circ, 11.635^\circ, 9.431^\circ$ and $3.890^\circ$. Though our numerical method allows to work with arbitrary values of the twist angle, these four values are chosen, close to the commensurate angles, to verify rigorously the symmetrical property of the conductivity tensor. We have verified the conductivity tensor has the following symmetry properties:
%
%
\begin{subequations}\label{symmetryTBG}
\begin{align}
\sigma_{xx}(\omega) &= \sigma_{yy}(\omega),\\
\sigma_{xy}(\omega) &= -\sigma_{yx}(\omega)\neq 0.
\end{align}
\end{subequations}
%
%
Additionally, the value of these elements are independent of the reference frame fixed for the calculation. In Fig.~\ref{fig_1}b) we show the optical conductivity tensors for several SBG configurations with different sliding vectors $\boldsymbol{\tau}$ with the length $\ell_{\boldsymbol{\tau}} = 0.8a_\mathrm{cc}, 0.6a_\mathrm{cc}, 0.4a_\mathrm{cc}, 0.2a_\mathrm{cc}$ and the angle $\phi_{\boldsymbol{\tau}} = 12^\circ$, where $a_\mathrm{cc}\approx 0.145$~nm is the nearest distance between two carbon atoms in the graphene monolayer. For SBGs, since the translational symmetry of the lattices is preserved, we calculated the optical conductivity tensor using the following two methods: the Kubo-Bastin formula in the real-space approach and the Kubo-Greenwood formula in the reciprocal lattice space approach (see~\ref{appA}). For the latter case, we express the conductivity tensor as
%
%
\begin{equation*}
\sigma_{\alpha\beta}(\omega)=\sum_{\mathbf{k}\in\mathrm{ BZ}}\sigma_{\alpha\beta}(\mathbf{k},\omega),
\end{equation*}
%
%
where the vector $\mathbf{k}$ is defined in the first Brillouin zone (BZ). We verified that the results from the two methods coincide. For SBGs, we found in general that conductivity tensor has the following symmetry properties:
%
%
\begin{subequations}\label{symmetrySBG}
\begin{align}
\sigma_{xx}(\omega)& \neq\sigma_{yy}(\omega), \\
\sigma_{xy}(\omega)& = \sigma_{yx}(\omega). 
\end{align}
\end{subequations}
%
%
This means that the SBG is optically anisotropic. The symmetry property is different from the case of TBGs in Eqs.~(\ref{symmetryTBG}). Additionally, the specific values of the conductivity tensor components depend on the choice of the Cartesian axes. However, the values for the conductivity tensor in two different Cartesian frames are related by the standard coordinate transformation $\sigma^\prime(\omega) = R_\varphi\sigma(\omega)R_\varphi^{-1}$, where $R_\varphi$ is the $2\times 2$ rotation matrix transforming one frame to the other. In particularly, we found that when the sliding vector $\boldsymbol{\tau}$ is either collinear or perpendicular to one of the vectors $\boldsymbol{\delta}_i$ with $i=1,2,3$, {\it i.e.} the vectors connecting one carbon atom to its three nearest neighbors in the honeycomb lattice,  the optical Hall conductivity $\sigma_{xy}(\omega)$ is zero in the reference frame with $\boldsymbol{\tau}$ collinear with the $Ox$ axis. 

These results for TBGs and SBGs are completely different from those for the AA- and AB-stacked configurations where the conductivity tensor is isotropic. In general, the appearance of a finite  optical Hall conductivity, and the relations between the tensor components, are not related to the breaking of TRS, but to the spatial symmetries of the atomic lattices. For the bilayer system, the AA-stacking configuration presents the highest symmetry with the point group $D_{6h}$ and the space group $p6mm$. The symmetry of the AB-stacked configuration is lower with the point group $D_{3d}$ and the space group $p3m1$. Introducing a finite value of the twist angle $\theta$ and the sliding vector $\boldsymbol{\tau}$ significantly reduces the symmetry of the resulting bilayer lattices. 
%
%
\begin{figure*}\centering
\includegraphics[width = \textwidth]{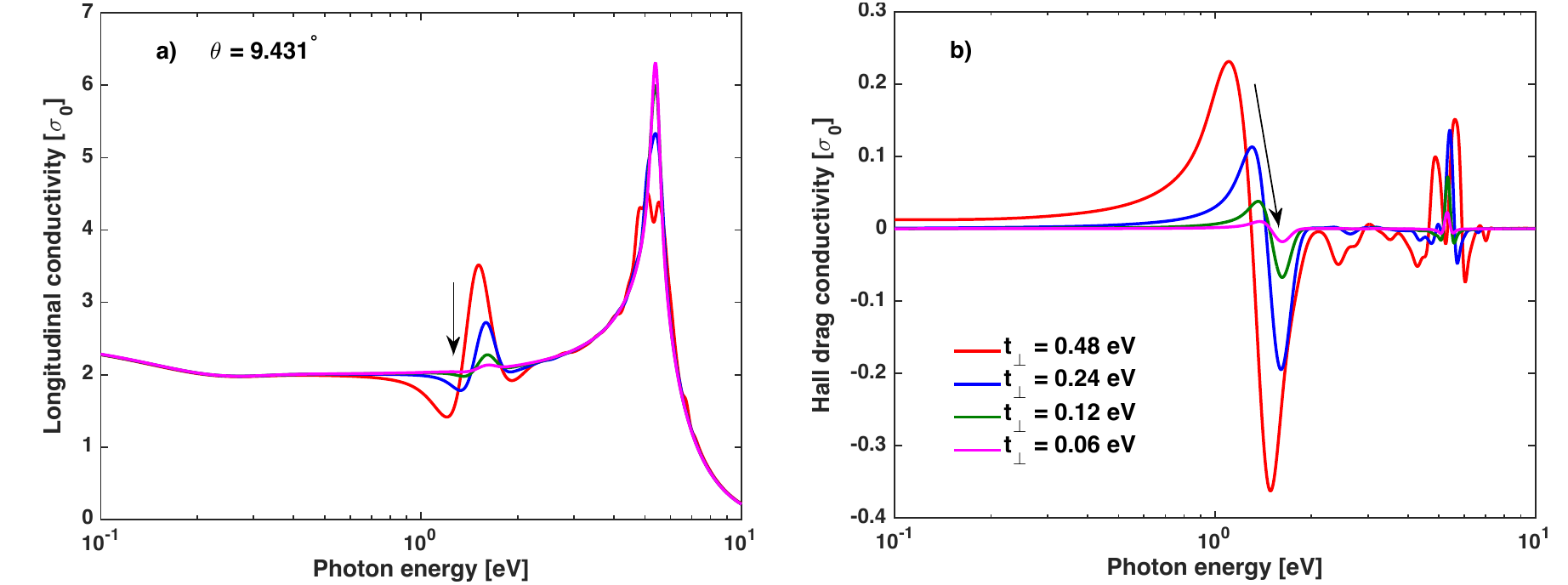}
\caption{\label{fig_2}The longitudinal [Panel a)] and transversal optical Hall [Panel b)] conductivities of a TBG configuration with $\theta = 9.431^\circ$ as a function of the photon energy for various values of the interlayer coupling parameter $t_\perp$. It shows the isotropy of the optical conductivity tensor in the limit of interlayer decoupling, $t_\perp\rightarrow 0$.}
\end{figure*}
%
%
Specifically, $\theta$ breaks the translational and mirror symmetries, thus reducing the point group of the TBG lattices to be $D_6$ (or $D_3$ depending on the position of the twist axis)~\cite{Zou-2018}. On the other hand, $\boldsymbol{\tau}$ breaks all point group symmetries, but it preserves the translational symmetry. However, when the sliding vector $\boldsymbol{\tau}$ is collinear or perpendicular to one of the three vectors $\boldsymbol{\delta}_i$, an axis $C_2^\prime$ exchanging the two layers and a mirror plane perpendicular to this rotation axis is preserved. 
These elements, together with an inversion center $\mathcal{I}$, form the point group $C_{2h}$. Within these symmetry considerations, we verify that both the TBG and SBG lattices are chiral. In fact, the TBG configurations with the twist angles of $\theta$ and $-\theta$ are the mirror images of each other, but they are never coincident. Similarly, the SBG configurations with $\boldsymbol{\tau}=(\tau_x,\tau_y)$ and $\boldsymbol{\tau}^\prime = (\tau_x,-\tau_y)$ are also the mirror images of each other and they are never identical if the lattice has no the mirror symmetry. Such point groups of the TBG and SBG lattices are given in the three-dimensional space. However, because of their 2D nature, the physical properties of these systems are governed by their 2D sub-groups, \emph{i.e.}, $C_s$ for SBGs and $C_6$ (or $C_3$) for TBGs. Thus, it is easy to verify that $\sigma_{xx}=\sigma_{yy}$ and $\sigma_{yx} = -\sigma_{xy}$ for TBGs and $\sigma_{xy} = \sigma_{yx}$ and $\sigma_{xx}\neq\sigma_{yy}$ for SBGs, confirming the data we have obtained numerically. The vanishing of $\sigma_{xy}(\omega)$ in special bilayer configurations, such as the SBG configurations with the $C_{2h}$ symmetrical point group and also the AA- and AB-stacked configurations, is clearly due to the cancelling contribution of optical transitions enforced by the mirror symmetry. 
Indeed, because of the preservation of a mirror plane in the SBGs with $\boldsymbol{\tau}\propto \boldsymbol{\delta}_i$, the Hamiltonian is even with respect to $k_y$, \emph{i.e.}, $\hat{\mathcal{H}}(k_x,k_y) = \hat{\mathcal{H}}(k_x,-k_y)$ but the electric current component $\hat{j}_y$ is odd since $\hat{v}_y(k_x,k_y) = (1/\hbar)\partial \hat{\mathcal{H}}(k_x,k_y)/\partial k_y$. As a consequence, the quantity $\sigma_{xy}(k_x,k_y)$ becomes odd with respect to $k_y$, \emph{i.e.}, $\sigma_{xy}(k_x,k_y) = -\sigma_{xy}(k_x,-k_y)$. As a result, the contribution from all Bloch states in the first BZ will mutually cancel leading to $\sigma_{xy}(\omega) = 0$. 
%
%
\begin{figure}
\centering
\includegraphics[width = \columnwidth]{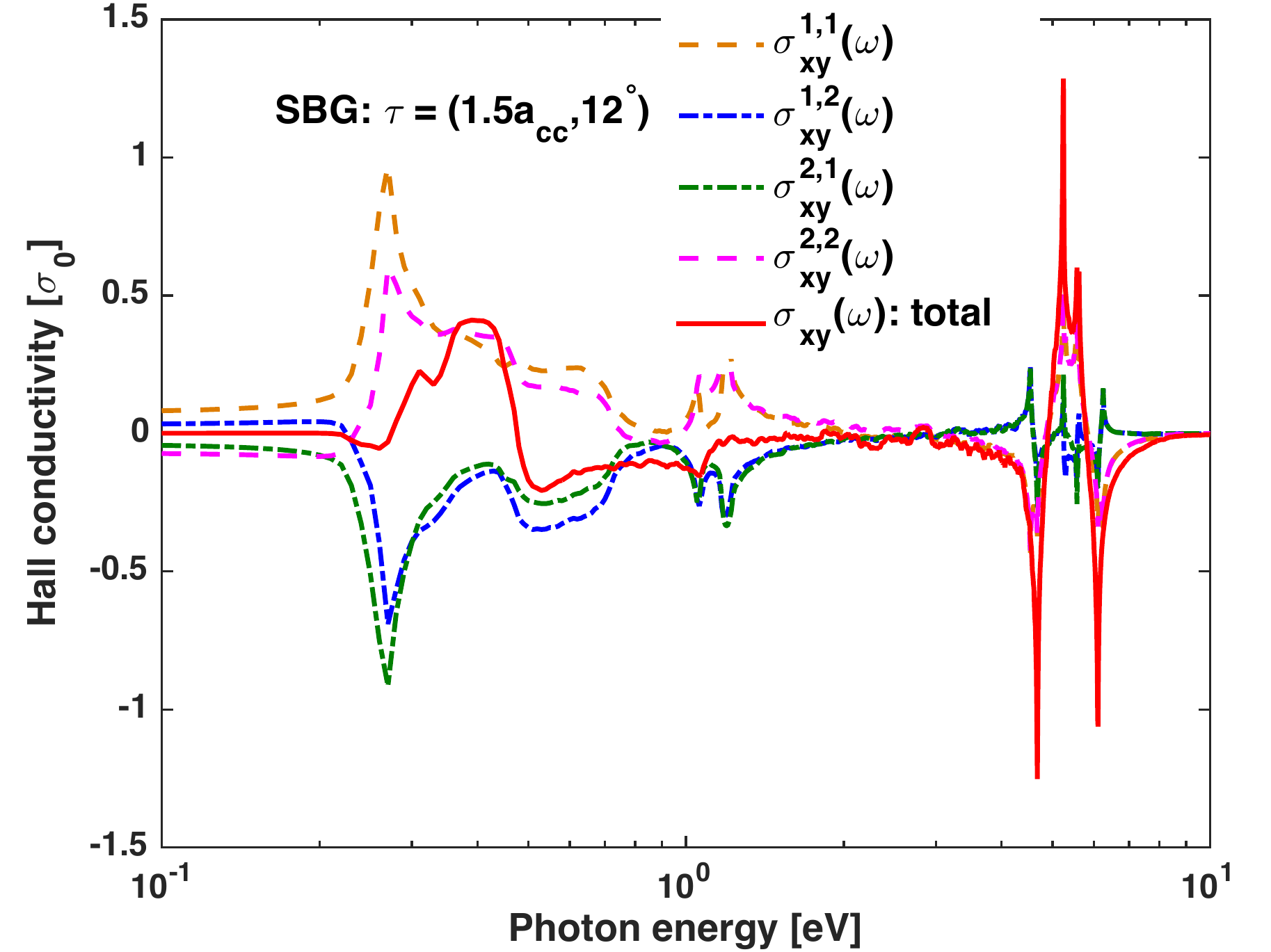}
\caption{\label{fig_3}The optical Hall conductivity $\sigma_{xy}(\omega)$ (the red solid curve) as the result of the correlation of various terms of the total velocity operators $\hat{v}_\alpha = \sum_\mu\hat{v}_\alpha^\mu$, $\mu = 1,2,12$. Here  $\sigma_{xy}^{\mu\nu}(\omega) \propto \langle \hat{v}_x^\mu(\omega)\hat{v}_y^\nu\rangle$. The data is presented for a SBG lattice with $\boldsymbol{\tau} = (1.5a_\mathrm{cc},12^\circ)$. The terms corresponding to the correlation of the interlayer current $\hat{\mathbf{v}}^{12}$ are small and not shown.}
\end{figure}
%
%
For TBGs, the interpretation of its optical activity is more subtle: Su\'arez Morell~{\it et al.} addressed it in terms of the rotation of the isospin of the graphene Weyl fermions~\cite{Morell-2017}. However, this is not sufficient because if the two graphene layers are decoupled, the behavior of the system must be identical to that of the monolayer, \emph{i.e.}, with $\sigma_{xy}(\omega)=0$. We can show that by decreasing the interlayer hopping parameter $t_\perp$, the longitudinal conductivity of TBGs approaches the value of twice the conductivity of monolayer graphene, and the Hall conductivity vanishes as seen in Figs.~\ref{fig_2}a) and~\ref{fig_2}b). Following  Su\'arez Morell~\emph{et al.}, we also decomposed the electron velocity operator $\hat{v}_\alpha$ into the terms involving the electron motion in each graphene layer, the in-plane or intra-layer velocities $\hat{v}_\alpha^{1(2)}$, and the inter-layer velocities $\hat{v}^{12}_\alpha$, \emph{i.e.}, $\hat{v}_\alpha = \hat{v}_{\alpha}^1+\hat{v}_{\alpha}^2+\hat{v}_{\alpha}^{12}$. By denoting $\langle \hat{v}_x^\mu(\omega) \hat{v}_y^\nu\rangle$ as the velocity-velocity correlation functions, according to the linear-response theory, we can assign the optical Hall conductivity $\sigma_{xy}^{\mu\nu}(\omega)$ to  $\sigma_{xy}^{\mu,\nu}(\omega) = ie^2\langle \hat{v}_x^\mu(\omega) \hat{v}_y^\nu\rangle/\omega$. Here we denote $\mu,\nu$ as the indices for the electron velocity terms, which take the value $\mu,\nu = 1,2,$ and $12$. From the Hamiltonian (\ref{Eq1}) these velocity terms are determined by:
%
%
\begin{subequations}
\begin{align}
    \mathbf{v}^\nu&=\frac{i}{\hbar}\sum_{i,j}t^m_{ij}(\mathbf{r}_j^\nu-\mathbf{r}_i^\nu)|\nu i\rangle\langle \nu j|~~\nu=\{1,2\}, \label{interJ} \\
    \mathbf{v}^{12}&=\frac{i}{\hbar}\sum_{\nu\neq\mu=1}^2\sum_{i,j}t^{\nu\mu}_{ij}(\mathbf{r}_j^\nu-\mathbf{r}_i^\mu)|\nu i\rangle\langle \mu j|, \label{intraJ}
\end{align}
\end{subequations}
%
%
the latter can be further simplified by decomposing $\mathbf{r}_j^2-\mathbf{r}_i^1=\mathbf{d}_\mathrm{GG}+\mathbf{r}_{ij}$, \emph{i.e.} into a vertical and a horizontal contribution, respectively. Here $\mathbf{d}_\mathrm{GG}$ is the vector vertically connecting the two graphene layers with the length $d_\mathrm{GG} = 0.335$ nm. The velocity in Eq.~(\ref{intraJ}),  therefore, can be expressed as the sum of two perpendicular contributions $\hat{\mathbf{v}}^{12}=\hat{\mathbf{v}}^{12}_z+\hat{\mathbf{v}}^{12}_\mathrm{drag}$. Since $\hat{\mathbf{v}}^{12}_\mathrm{drag}$ lies in the lattice plane, only this component contributes to the velocity-velocity correlator.  

In Fig.~\ref{fig_3} we display data for a SBG configuration with the sliding vector $\boldsymbol{\tau} = (1.5a_\mathrm{cc},12^\circ)$. We see that the magnitude of $\sigma_{xy}^{1,1}(\omega)$ and $\sigma_{xy}^{2,2}(\omega)$ are comparable to those of $\sigma_{xy}^{1,2}(\omega)$ and $\sigma_{xy}^{2,1}(\omega)$, while those of $\sigma_{xy}^{12,12}(\omega)$ and $\sigma_{xy}^{1(2),12}(\omega)$ are negligible and not shown. These data indicate clearly that the appearance of the optical Hall conductivity is not dictated solely by the correlation of the electron velocities in two different graphene layers, but by the correlation of the velocities in the same graphene layer as well. A different explanation was proposed by Kim~{\it et al.} in Ref.~\cite{Kim-2016}. They stated that the circular dichroism of TBGs is due to the interlayer optical transitions. However, the interlayer optical transitions  occur as well in the AA- and AB-stacked configurations but $\sigma_{xy}(\omega) = 0$. All these analyses suggest that we need to pay particular attention to determining the essential factors governing the optical transitions, and hence, the velocity-velocity correlation: the electronic states conducting the current. Unfortunately, it is simply impossible to visualize these states. However, in the following, we will present a way to analyze their behavior through the dynamics of wave packets.

\section{Wave-packet dynamics}\label{wpd}
%
%
\begin{figure*}\centering
\includegraphics[width=\textwidth]{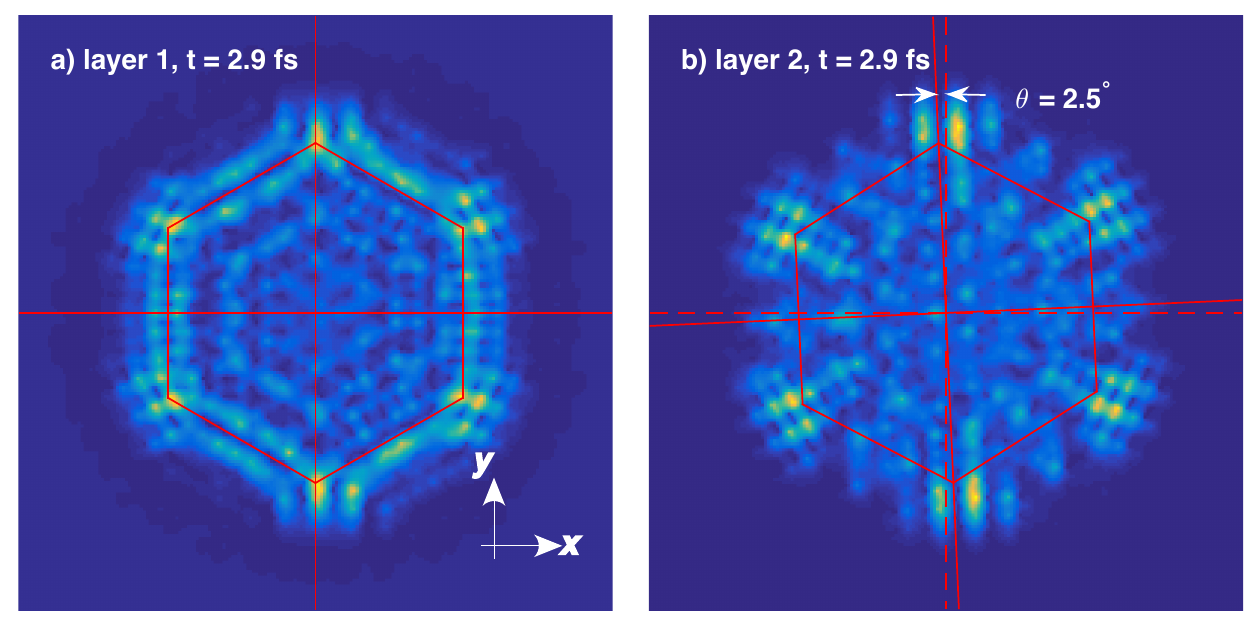}
\caption{\label{fig_4} The snapshot at $t = 2.9$ fs of the distribution of the probability density of an electron in the graphene layer 1 [a)]/layer 2 [b)] of a TBG lattice with $\theta = 2.5^\circ$. }
\end{figure*}
%
%
%
%
\begin{figure*}\centering
\includegraphics[width=\textwidth]{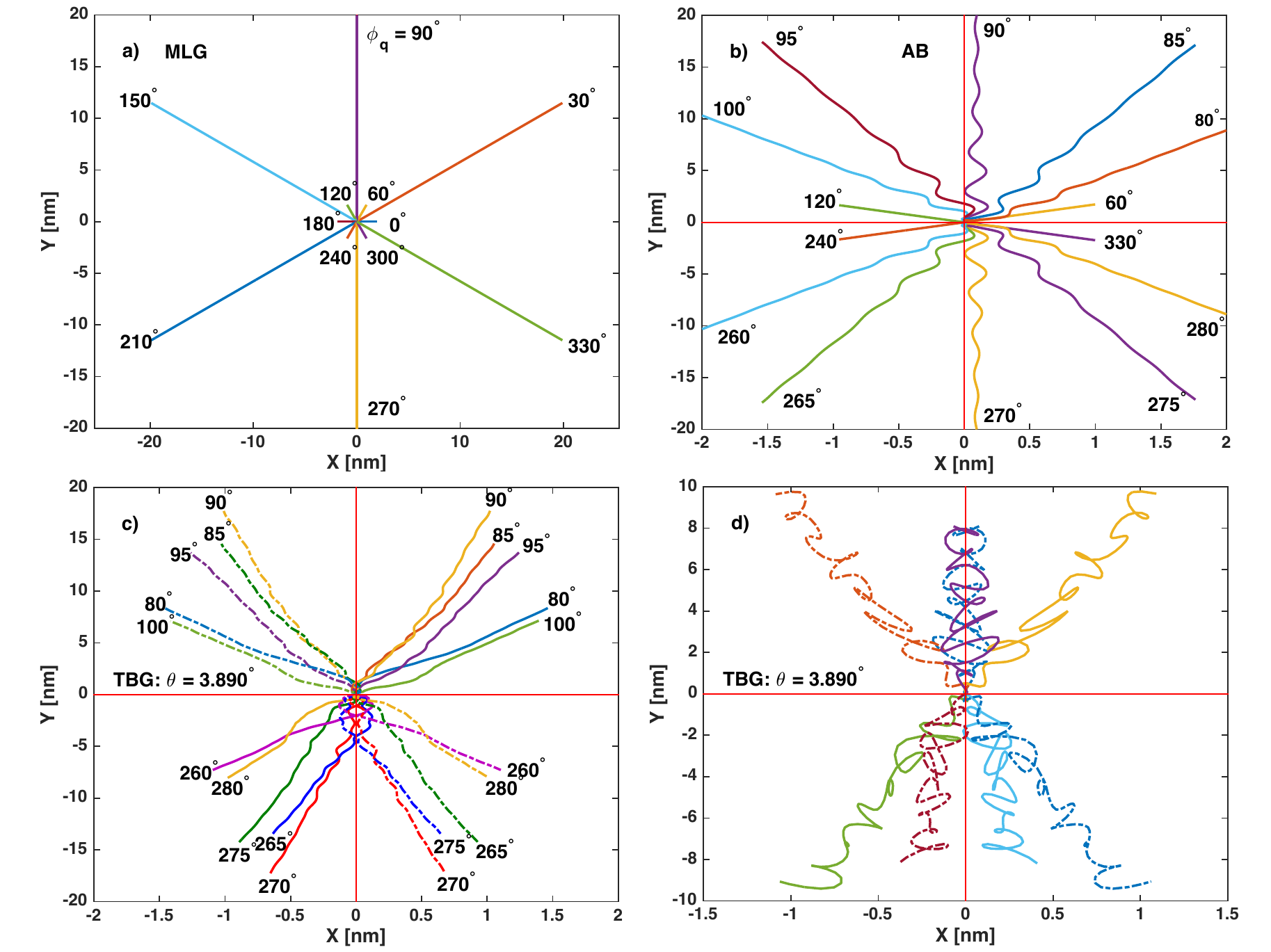}
\caption{\label{fig_5}The trajectory of kicked Gaussian wavepackets with $\xi = 2\sqrt{3}a_{cc}$ and the initial wave vector of $q = 0.95\times \pi/3\sqrt{3}a_\mathrm{cc}$ orienting along various directions characterized by the angle $\phi_q$: a) in the monolayer graphene, b) in the AB-stacked bilayer graphene, c) in a TBG graphene with the twist angle $\theta = 3.890^\circ$. In Panels~a) and~b) the zigzag lines of the honeycomb lattice orient along the directions of $30^\circ, 90^\circ$ and $150^\circ$, showing the preferable directions of electron propagation. In the Panel~c) the solid/dashed curves are for the wave packets that are initially located in the top/bottom graphene layer. The Panel~d) shows the deviation from each other of the trajectories of the parts of the Gaussian wavepackets propagating on the top (solid)/bottom (dashed) graphene layer of a TBG lattice for $\phi_{q} = 90^\circ$ and $270^\circ$.}
\end{figure*}
%
%
To unveil the physics of the optical Hall response, we numerically tracked the time evolution of electrons in  TBG lattices. In Fig.~\ref{fig_4} we display a snapshot at time 2.9~fs of the distribution of the probability density of an electron initially occupying a single $2p_z$ orbital in layer~1. We observed that the electron wave does not spread solely in layer~1, but it penetrates and spreads into layer~2 as well. The electron wave propagation is always interchangeable between the two layers. This implies the existence of hybridized states that supports such a wave interchange. Noticeably, the wavefronts of the electron waves in two layers present differences and similarities: both wavefronts show the anisotropy of the wave spreading along the six preferable directions parallel to the zigzag lines of the honeycomb lattice~\cite{Do-2019}. However, the relative rotation of the two lattices shows the misalignment of the preferable directions of electron propagation in  layer~2 compared to  layer~1. This result partially supports the conclusion by Su\'arez Morell~\emph{et al.} in Ref.~\cite{Morell-2017}. To clarify the key role played by the hybridized states in governing the finite Hall conductivity, we investigated the evolution of kicked Gaussian waves 
%
%
\begin{equation}
\psi(\mathbf{r},t=0) \propto \exp\left[-\frac{(\mathbf{r}-\mathbf{r}_0)^2}{4\xi^2}\right]\exp(i\mathbf{qr}),
\end{equation}
%
%
where $\mathbf{r}_0$ is the initial center of the wave packet, $\xi$ is its width and $\mathbf{q}$ is the initial wave vector. 
As a first step, we tracked the trajectory of the wave centroid in the lattice of monolayer graphene. The wave centroid at time $t$ is defined by the vector 
%
%
\begin{equation}
\mathbf{r}_\mathrm{c}(t) = \sum_{i}\mathbf{r}_i|\psi(\mathbf{r}_i,t)|^2 
\end{equation}
%
%
where the summation is over all lattice nodes $\mathbf{r}_i$. We observed that the wave centroid always evolves along straight lines parallel to the direction of the initial wave vector $\mathbf{q}$ independently of the zigzag and armchair directions of the honeycomb lattice | see Fig.~\ref{fig_5}a). Semi-classically, this implies that an electron injected into the honeycomb lattice with an initial velocity $\mathbf{v}_\mathbf{q}(0)\propto \mathbf{q}$ will move along this direction without any deflection, \emph{i.e.}, $\mathbf{v}_\mathbf{q}(t) \propto \mathbf{q}$ at time $t > 0$. Denoting by $v_{q_\|}$ and $v_{q_\perp}$ the components of the velocity $\mathbf{v}_\mathbf{q}$ of the wave centroid parallel and perpendicular to $\mathbf{q}$, respectively, we have $v_{q_\|} = v_\mathbf{q}$ and $v_{q_\perp} = 0$. This yields $\langle v_{q_\perp}(\omega)v_{q_\parallel}\rangle = 0$. Since $\sigma_{xy}(\omega)$ can be regarded as the result of the average of the velocity-velocity correlation functions over all possible values of $\mathbf{q}$ and $\mathbf{r}_0$, \emph{i.e.}, $\sigma_{xy}(\omega)\propto  \overline{\langle v_{q_\perp}(\omega)v_{q_\parallel}\rangle}$, it therefore explains  the zero optical Hall conductivity of monolayer graphene. We note in passing that the same argument applies for the case of AA-stacked bilayer graphene. However, for the AB-stacked lattice, we observed the oscillation behavior of the wave centroid trajectories along the six preferable directions of the electron propagation | see Fig.~\ref{fig_5}b). By decreasing the interlayer hopping parameter $t_\perp$, the oscillation amplitude of the trajectories is reduced. The oscillation trajectories are a peculiar feature of the hybridized electron states in the AB-stacked lattice. Remember that in this atomic lattice, three $\sigma_v$ mirror planes of the honeycomb lattice are broken, but they replaced by three $C_2^\prime$ axes that interchange the two graphene layers. The motion of an electron is thus not constrained by  mirror symmetry, but by the $C_2^\prime$ symmetry. More importantly, the oscillation trajectories indicate that an electron gains a nonzero transverse velocity $v_{q_\perp}\ne 0$ when moving in the lattice, leading to  $\langle v_{q_\perp}(\omega)v_{q_\|}\rangle \ne 0$. However, analyzing in details Fig.~\ref{fig_5}b) we can infer the zero Hall conductivity $\sigma_{xy}(\omega) = 0$ by noticing that there are always two mirror symmetric trajectories related by the symmetry plane $\sigma_d$ of the AB-stacked lattice corresponding to two distinct values of $\mathbf{q}$. We note in passing that a similar behavior is also observed for the particular SBG  configurations with the $C_{2h}$ symmetry. This suggests that the existence of a mirror symmetry plane in the bilayer lattices will always lead to the existence of pairs of momenta $\mathbf{q}$ and $\mathbf{q}'$ such that  $v_{q'_\|} = -v_{q_\|}$, but $v_{q'_\perp} =  v_{q_\perp}$. As a consequence, these terms always cancel each other on average,  resulting in the zero optical Hall conductivity, {\it i.e.}, $\sigma_{xy}(\omega) = 0$.  In Fig.~\ref{fig_5}c) we show the trajectories of the wave centroid of kicked Gaussian wavepackets in a TBG lattice. The solid (dashed) curves are for the cases in which the initial wave packets are located in layer 1 (layer 2). We clearly see the deflection of the trajectories from the lines along the initial vector $\mathbf{q}$, which means that $\langle v_{q_\perp}(\omega)v_{q_\|}\rangle \ne 0$. Because of the absence of the mirror symmetry, the TBG lattices are chiral. This dictates as well the chirality of the hybridized electron states as the mirror images of the solid and dashed curves shown in Fig.~\ref{fig_5}c). A further study of the centroids of the wave parts propagating on two graphene layers shows that they moves along the different curly trajectories | see Fig.~\ref{fig_5}d). This explains  the deflection of the trajectories and the left-, right-deflection (chirality) behaviors observed in Fig.~\ref{fig_5}c). So, with the same argument made for the monolayer and AB-stacked bilayer we conclude that the TBG lattices will be characterized by  finite optical Hall conductivity since there are no cancellation contributions to $\langle v_{q_\perp}(\omega)v_{q_\|}\rangle$ in the optical Hall conductivity due to the breaking of the mirror symmetry, \emph{i.e.}, $\sigma_{xy}(\omega)\propto \overline{\langle v_{q_\perp}(\omega)v_{q_\|}\rangle}\neq 0$ after averaging over $\mathbf{q}$ and $\mathbf{r}_0$.

\section{Faraday, Kerr rotation and Circular dichroism}\label{secIV}

\begin{figure}\centering
\includegraphics[width=\columnwidth]{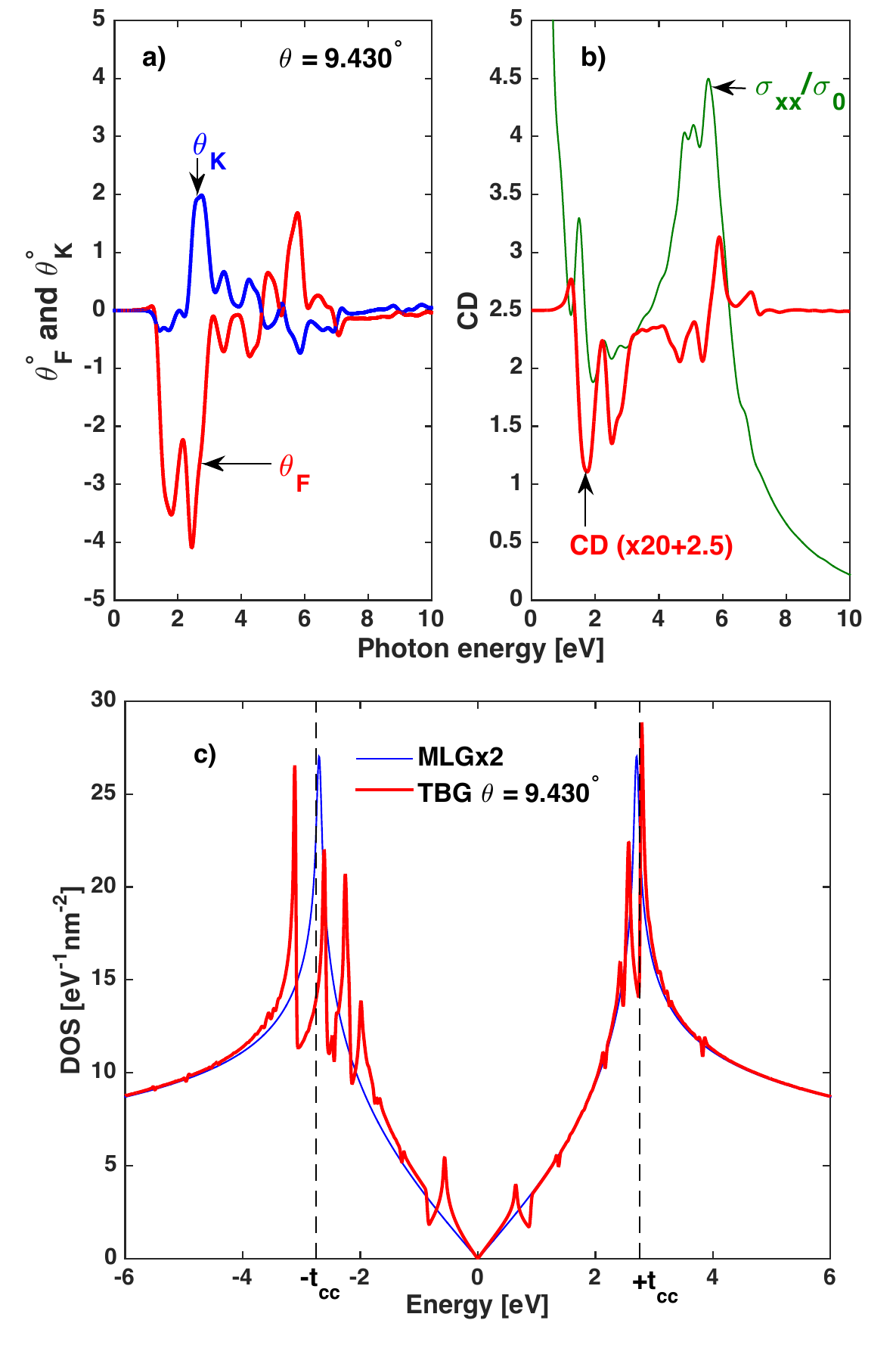}
\caption{\label{fig_6}(a) The Faraday $\theta_\mathrm{F}$ and Kerr $\theta_\mathrm{K}$ rotation angles and (b) the circular dichroism as a function of the photon energy for a representative TBG configuration with the twist angle $\theta=9.430^\circ$ versus the photon energy. The plot of the CD is shifted upward by an amount $2.5$, and the data are multiplied by $20$ to compare with $\sigma_{xx}$. The CD is displayed (the red curve) together with the real part of the conductivity $\sigma_{xx}/\sigma_0$ (the blue curve) to associate the structure of the former with that of the conductivity. (c) The DOS as a function of the energy for  TBG with $\theta = 9.430^\circ$ in red. The DOS of the totally decoupled BLG (in blue) is also displayed to highlight the energy ranges in which the hybridized states are manifested through the sub-peaks of the red curve.} 
\end{figure}

In the previous section we saw that the electrical conductivity tensor is the key quantity to characterize the transport and optical properties of an electronic system. To complete our discussion, we present in this section results for the Faraday and Kerr rotation angles of the light polarization vector as well as the CD, a quantity quantifying the difference of the absorption of the left-handed and right-handed circular polarization light. We employed the transfer matrix method to determine the transmission $\mathbf{t}$ and reflection $\mathbf{r}$ matrices; these express the relationship between the amplitude of the transmitted/reflected light and that of the incident light~\cite{Szechenyi-2016}. The details of the calculation are presented in Appendix~\ref{appC} where the relationship between these matrices and the components of the electrical conductivity tensor is presented. In particular, we show that $t_{xy}(\omega), r_{xy}(\omega) \propto \sigma_{xy}(\omega)$, see Eqs.~(\ref{B25}) and~(\ref{B26}). From these results, together with Eqs.~(\ref{B18}) and~(\ref{B13}), we see that there are no Faraday or Kerr rotation nor any CD if the systems do not have the Hall response, $\sigma_{xy}(\omega)\neq0$. For the case of SBG systems, because of the specific symmetry properties of the conductivity tensor, \emph{i.e.}, $\sigma_{xy}(\omega)=\sigma_{yx}(\omega)$, Eq.~(\ref{B32}) indicates that there is no CD. In other words, the SBG systems cannot distinguish the left-handed and right-handed circular polarized light despite the chirality of the atomic lattice. This is different from the case of TBG systems since $\sigma_{xy}(\omega)=-\sigma_{yx}(\omega)$. In Fig.~\ref{fig_6}a) and Fig.~\ref{fig_6}b) we present the values of the Faraday, and Kerr rotation angles and of the CD calculated for a TBG configuration with the twist angle $\theta=9.430^\circ$ (The results for the other twist angles are qualitatively similar).
%
%
%
%

These results show that these quantities vary versus the photon energy $\omega$ of the incident light. In the low ($< 1$~eV) and high ($> 7$~eV) energy ranges wherein the electronic states in two graphene layers effectively decouple, the values of $\theta_\mathrm{F}, \theta_\mathrm{K}$ and the CD are zero as expected. On the contrary, in the energy range of $(1,7)$~eV where the hybridized electron states are formed and manifest as the peaks in the density of states~\cite{Le-2018},
the values of $\theta_\mathrm{K},\theta_\mathrm{F}$ and CD are different from zero. For the TBG configuration with $\theta=9.430^\circ$ we observe that $\theta_\mathrm{F}$ can reach a value of $4^\circ$, while $\theta_\mathrm{K}$ can reach a value  of $2^\circ$, and the CD is $\sim 8 \%$. In general, the dependence of these quantities on the photon energy $\omega$ is complicated, as is presented in Eq.~(\ref{B32}). Physically, since the CD is associated with the light absorption, the behavior of the curve $\mathrm{CD}(\omega)$ should relate to the real part of the longitudinal conductivities $\sigma_{xx}(\omega)$ and $\sigma_{yy}(\omega)$. In Fig.~\ref{fig_6}b) we present the curve $\mathrm{CD}(\omega)$ together with the curve $\sigma_{xx}(\omega)$ in which the CD values are multiplied by 20  and shifted upward by an amount of 2.5 as a guide for the eyes. Clearly, we observe the consistency of the behavior of the $\mathrm{CD}(\omega)$ result with that of the conductivity curve $\sigma_{xx}(\omega)$.

\section{Discussion and conclusions}\label{secV}
Before concluding the paper we would like to validate the available predictions of the physical properties of generic TBG systems that were usually deduced for commensurate configurations. As our calculation method is based on the real-space approach, it can be applied to lattices of arbitrary stacking, regardless of commensurate or incommensurate configuration. Using our numerical method, we can conclude that both the density of states and the conductivity tensor $\boldsymbol{\sigma}$ vary continuously  with the twist angle $\theta$ and the sliding vector $\boldsymbol{\tau}$ in the whole range of these parameters, \emph{i.e.}, $\theta\in (0^\circ,60^\circ)$ and $\boldsymbol{\tau}$ given in the triangle defined by two unit vectors $\mathbf{a}_{1}$ 
and $\mathbf{a}_{2}$ of the honeycomb lattice. However, it is worth noticing that the behaviors of the AA- and AB-stacked configurations cannot be deduced as limiting cases of the TBG system for $\theta \rightarrow 0^\circ$ or $60^\circ$ or of the SBG system for $\boldsymbol{\tau}\rightarrow 0$. We confirm this argument by a symmetry analysis: as long as $\theta\neq 0$ or $60^\circ$ or $\boldsymbol{\tau}\neq 0$ the system symmetries are not changed by varying these parameters. A sudden change is obtained only when either $\theta$ or $\boldsymbol{\tau}$ are equal to the limit values. For these cases, the point group $D_6$ (or $D_3$) of the TBG lattices changes to $D_{6h}$  (or to $D_{3d}$) of the AA-stacked lattice (or of the AB-stacked lattice), whereas for the case of SBG, the change is from $E$ to $C_s$ or to higher symmetry point groups. Additionally, for the TBG case, there will be a collapse of the unit cell from the very large size to the one containing only four carbon atoms when $\theta$ changes to $0^\circ$ or $60^\circ$. In both the AA- and AB-stacked lattices, the number of carbon atoms in one layer coupled to another carbon atom in the second layer is higher than that in the TBG and SBG lattices. Indeed, by defining $\bar{n}_{nb}$ as the average number of lattice nodes on one layer that electron can hop from another node in the other layer, we obtain the following:  for a given radius of proximity $r_c = \sqrt{a_\mathrm{cc}^2+d^2}$, we find that it is $\bar{n}_{nb} = 4(5)$  for $\theta = 0^\circ(60^\circ)$, but $\bar{n}_{nb} \approx 2.7$ for  different values of $\theta$, regardless of the size of $\theta$. This observation is interesting because it can help to explain the behavior of the effective decoupling of the two layers in some energy ranges~\cite{Sboychakov-2018,Le-2018}.

In conclusion, stacking material layers of atomic thickness has been considered as a potential path for engineering the electronic structure and physical properties of complex 2D material systems, especially when attempting to design  devices for manipulating light at the nanoscale~\cite{bao_2019,Akinwande_2019}. This may be of potential interest for designing optical systems that can to distinguish between different enantiomers of molecules, with important application in medicine and chemistry~\cite{Izake_2007,Liu_2005,Solomon_2019}. In this work, we analyzed in-depth the origin of the finite optical Hall response of bilayer graphene under twisting and/or sliding  two layers. We showed that lattices of twisted- and slid-bilayer graphene are chiral, and they can be used to help rotate the polarization of linearly polarized light. Our analysis was based on a real-space computation scheme developed to compute all the components of the optical conductivity tensor. We showed in detail that the TBG lattices are isotropic and support the CD, while the SBG lattices are anisotropic and do not support the CD. The calculation method allows us to monitor the evolution of an electron in atomic bilayer lattices with an arbitrary twist angle and sliding vector. The chiral behavior of hybridized electron states was determined to be a deflection of the trajectory of the kicked Gaussian wave packets in  BLG lattices. The optical Hall response of the BLG system, therefore, was argued to be a manifestation of the chirality of the hybridized states that supports the interchange of electrons between the two graphene layers. However, we showed that the mirror symmetry constrains the contribution of such states to the optical Hall response. In the lattices without mirror symmetry, such as the TBGs and SBGs, the hybridized states govern the correlation of different components of the electron velocity in  such a way that the terms do not cancel each other, hence resulting in nonzero optical Hall conductivity. To quantify the optical activity of the bilayer graphene systems we employed the transfer matrix method to establish the relations of the transmission and reflection matrices to the components of the conductivity tensors, and then we determined the Faraday and Kerr rotation angles as well as the circular dichroism. Finally, taking advantage of the calculation method, combined with a symmetry analysis, we concluded that there is a continuous variation of physical quantities, including the density of states and the electrical conductivity tensors, on the twist angle and the sliding vector. This conclusion can be applied to bilayer graphene systems that would be deduced using the force-brute  exact diagonalization approach.

\section{Acknowledgements}
We acknowledge discussions with A.~Garc\'ia-Etxarri and T.~van~den~Berg.
The work is supported by the National Foundation for Science and Technology Development (NAFOSTED) under Project No. 103.01-2016.62. The work of DB is supported by Spanish Ministerio de Ciencia, Innovation y Universidades (MICINN) under the project FIS2017-82804-P, and by the Transnational Common Laboratory \textit{Quantum-ChemPhys}. 

\appendix
\section{The Kubo-Bastin formula}\label{appA}

There are a number of versions of the Kubo formula for  electrical conductivity that can be implemented in different situations. Here we present the derivation for Eq.~\eqref{Eq1}. From the general linear response theory, the element $\sigma_{\alpha\beta}(\omega)$ of the electrical conductivity tensor is composed of  diamagnetic and paramagnetic components, $\sigma_{\alpha\beta}(\omega)= \sigma_{\alpha\beta}^A(\omega)+\sigma_{\alpha\beta}^P(\omega)$  where~\cite{Bruus-2016}:
%
%
\begin{subequations}
\begin{align}
\sigma^A_{\alpha\beta}(\omega) &= i\frac{e^2}{m_e\omega}n_e\delta_{\alpha\beta}\label{A1},\\
\sigma^P_{\alpha\beta}(\omega) &= \frac{ie^2}{\omega}\frac{1}{\Omega}\sum_{\ell,n}(f_n-f_\ell)\frac{\langle n|\hat{v}_\alpha|\ell\rangle\langle \ell|\hat{v}_\beta|n\rangle}{\hbar(\omega+i\eta)-(E_\ell-E_n)}\label{A2}
\end{align}
\end{subequations}
%
%
Here $m_e$ is the bare electron mass, $n_e$ is the electron density in a system, $\eta$ is a positive infinitesimal number, and $\Omega$ is the spatial volume of the considered system. The diamagnetic part is diagonal. It is determined through the calculation of $n_e$:
%
%
\begin{equation}\label{A3}
n_e = \int_{-\infty}^{+\infty}dE\rho(E)f(E) = \int_{-\infty}^{+\infty}dE\frac{\rho(E)}{1+e^{\beta(E-\mu)}},
\end{equation}
%
%
where $\rho(E)$ is the density of states of electrons, $\beta = 1/k_\text{B}T$ is the inversion of thermal energy, and $\mu$ is the chemical potential. For the paramagnetic part of the conductivity elements $\sigma^P_{\alpha\beta}(\omega)$, the properties of the delta-Dirac function involving the integration  are written as follows:
%
%
\begin{widetext}
\begin{align}\label{A4}
\sigma^P_{\alpha\beta}(\omega) &= \frac{ie^2}{\omega}\frac{1}{\Omega}\sum_{\ell,n}\int_{-\infty}^{+\infty}dE\delta(E-E_n)f(E)\frac{\langle n|\hat{v}_\alpha|\ell\rangle\langle \ell|\hat{v}_\beta|n\rangle}{E+\hbar\omega-E_\ell+i\hbar\eta}\nonumber\\
&+\frac{ie^2}{\omega}\frac{1}{\Omega}\sum_{\ell,n}\int_{-\infty}^{+\infty}dE\delta(E-E_\ell)f(E)\frac{\langle n|\hat{v}_\alpha|\ell\rangle\langle \ell|\hat{v}_\beta|n\rangle}{E-\hbar\omega-E_n-i\hbar\eta}
\end{align}
\end{widetext}
%
%
Now, noticing that $\delta(E-\hat{\mathcal{H}})|n\rangle = \delta(E-E_n)|n\rangle$ and introducing the retarded $(+)$ and advanced $(-)$ resolvents, we have:
%
%
\begin{align}
\hat{\mathcal{G}}^\pm(E\pm\hbar\omega) &= \frac{1}{E\pm(\hbar\omega+i\delta)-\hat{\mathcal{H}}}\label{A5}
\end{align}
%
%
Equation~\eqref{A4} is written in the form of Eq.~\eqref{Eq1}:
%
%
\begin{widetext}
\begin{equation}\label{A7}
\sigma^P_{\alpha\beta}(\omega) = \frac{ie^2}{\omega}\frac{1}{\Omega}\int_{-\infty}^{+\infty}dEf(E)\text{Tr}\left[\delta(E-\hat{\mathcal{H}})\hat{v}_\alpha\hat{\mathcal{G}}^+(E+\hbar\omega)\hat{v}_\beta+\hat{\mathcal{G}}^-(E-\hbar\omega)\hat{v}_\alpha\delta(E-\hat{\mathcal{H}})\hat{v}_\beta\right]
\end{equation}
\end{widetext}
%
%
For low frequencies we can approximate
%
%
\begin{equation}
    \mathcal{G}^\pm(E\pm\hbar\omega) \approx \mathcal{G}^\pm(E)\pm\frac{d \mathcal{G}^\pm(E)}{d E}\hbar\omega.
\end{equation}
%
%
So, we determine the real and imaginary parts of the conductivity $\sigma^P_{\alpha\beta}$ as follows
%
%
\begin{widetext}
\begin{subequations}
\begin{align}
    \text{Re}[\sigma^P_{\alpha\beta}(\omega)] &= -\frac{e^2\hbar}{\Omega}\int_{-\infty}^{+\infty}dEf(E)2\text{Im}\left(\text{Tr}\left[\delta(E-\hat{\mathcal{H}})\hat{v}_\alpha\frac{d\hat{\mathcal{G}}^+(E)}{dE}\hat{v}_\beta\right]\right),\\
    \text{Im}[\sigma^P_{\alpha\beta}(\omega)] &= +\frac{e^2}{\Omega}\frac{1}{\omega}\int_{-\infty}^{+\infty}dEf(E)2\text{Re}\left(\text{Tr}\left[\delta(E-\hat{\mathcal{H}})\hat{v}_\alpha\hat{\mathcal{G}}^+(E)\hat{v}_\beta\right]\right).
\end{align}
\end{subequations}
%
%
\end{widetext}
The imaginary part is inversely dependent on $\omega$ but the real part is independent of $\omega$. The real part is identical to the Kubo-Bastin formula that defines the dc conductivity~\cite{Bastin-1971}. As the delta-function and Green functions can be expanded efficiently in terms of Chebyshev polynomials, i.e., with the expansion coefficients given analytically, the Kubo-Bastin formula is useful for general calculation. In Ref.~\cite{Garcia-2015} the authors demonstrated successfully the calculation for the dc conductivity of topological systems.

\section{Retarded and advanced resolvents in terms of Chebyshev polynomials}\label{appB}
The Bessel function of the first kind is defined by the integral:
%
%
\begin{equation}
J_n(z) = \frac{1}{i^n\pi}\int_0^\pi d\theta \cos(n\theta)e^{iz\cos\theta}.
\end{equation}
%
%
This function has the property: $J_n(-z) = (-1)^nJ_n(z)$.

In terms of the Chebyshev polynomials of the first kind, it is straightforward to expand the exponent function $e^{\pm ixt}$. This yields
%
%
\begin{equation}
e^{\pm ixt}=\sum_{n=0}^{+\infty}\frac{2}{\delta_{n,0}+1}(\pm i)^n J_n(t)T_n(x).
\end{equation}
%
%
where $x \in (-1,1)$.

Applying this result to expand the retarded and advanced resolvents $\hat{\mathcal{G}}^\pm(E)=1/(E\pm i\eta-\hat{H}) = W^{-1}/(\epsilon\pm i\eta-\hat{h})$, where $\epsilon = (E-E_0)/W, \hat{h} = (\hat{H}-E_0)/W$, we obtain:
%
%
\begin{align}
\hat{\mathcal{G}}^\pm(E) &= \pm\frac{1}{iW}\int_0^{+\infty}dt e^{\pm i(\epsilon\pm i\eta)t}e^{\mp i\hat{h}t}\\
&=\pm \frac{1}{W}\sum_{n=0}^{+\infty}(\mp i)^{n+1}\frac{2}{\delta_{n,0}+1}g_n^\pm(\epsilon\pm i\eta)T_n(\hat{h})\nonumber
\end{align}
where $g_n^\pm(\epsilon\pm i\eta)$ are defined by: \cite{Ferreira-2015}
%
%
\begin{eqnarray}
g_n^\pm(z) &=& \int_0^{+\infty}dt e^{\pm izt}J_n(t)=\frac{(\sqrt{1-z^2}\pm iz)^n}{\sqrt{1-z^2}}
\end{eqnarray}
%
%
where $z= \epsilon\pm i\eta$. 

The derivative of the resolvents $\hat{\mathcal{G}}^\pm(E)$ with respect to $E$ is driven as follows:
%
%
\begin{align}
\frac{\partial\hat{\mathcal{G}}^\pm(E)}{\partial E} 
&=\pm\frac{1}{W}\sum_{n=0}^{+\infty}\frac{2}{\delta_{n,0}+1}(\mp i)^{n+1}\frac{\partial g_n^\pm(\epsilon\pm i\eta)}{\partial E}T_n(\hat{h})\nonumber\\
&=\pm\frac{1}{W^2}\sum_{n=0}^{+\infty}\frac{2}{\delta_{n,0}+1}(\mp i)^{n+1}\frac{\partial g_n^\pm(z)}{\partial z}T_n(\hat{h}),\nonumber
\end{align}
%
%
where $z = \epsilon\pm i\eta$ and
%
%
\begin{equation}
\frac{\partial g^\pm_n(z)}{\partial z} = \frac{1}{\sqrt{1-z^2}} \left(\frac{z}{\sqrt{1-z^2}}\pm in\right)g^\pm_n(z).
\end{equation}
%
%


\section{Faraday, Kerr rotation and circular dichroism}\label{appC}
\subsection{Elliptical polarization light}\label{appC1}
A monochromatic light is described by an electric vector with the components given in the form:
%
%
\begin{subequations}
\begin{align}
E_x(t,z) &= E_{0x}\cos(kz-\omega t)\label{B1}\\
E_y(t,z) &= E_{0y}\cos(kz-\omega t+\varphi)\label{B2}
\end{align}
\end{subequations}
%
%
where $\varphi$ is the dephasing between the two components $E_y$ and $E_x$. This light is elliptically polarized and is characterized by two parameters, \emph{i.e.}, the polarization angle $\alpha$ and the ellipticity $\tan\epsilon$ ($\epsilon$ is called the ellipticity angle). These two parameters are straightforwardly determined by:
%
%
\begin{subequations}
\begin{align}
\tan(2\alpha) &= \frac{2E_{0x}E_{0y}}{E_{0y}^2-E_{0x}^2}\cos\varphi,\label{B3}\\
\tan\epsilon &= \left| \frac{E_{0x}\tan\alpha+E_{0y}}{E_{0x}-E_{0y}\tan\alpha}\right|\label{B4}
\end{align}
\end{subequations}
%
%
where $\alpha\in(-\pi/4,\pi/4)$ and $\epsilon \in [0,\pi/2)$.

In practice, we usually use a complex field to represent a trigonometric function. We thus define a complex vector, named the Jones vector, for the electric field as follows:
%
%
\begin{equation}\label{B5}
\mathbf{E} = \frac{1}{\sqrt{E_x^2+E_y^2}}\begin{pmatrix}E_x\\ E_y\end{pmatrix} =  \frac{1}{\sqrt{E_{0x}^2+E_{0y}^2}}\begin{pmatrix}E_{0x}\\ E_{0y}e^{i\varphi}\end{pmatrix}\!.\!\!
\end{equation}
%
%
So, a monochromatic light of the vertical and horizontal linear polarization ($\varphi=0$) is defined by the following Jones vectors:
%
%
\begin{equation}\label{B6}
\tilde{\mathbf{E}}_v^\ell =\left(\begin{array}{c}0\\ 1\end{array}\right) \,\,\text{and}\,\, \tilde{\mathbf{E}}_h^\ell =\left(\begin{array}{c}1\\ 0\end{array}\right).
\end{equation}
%
%
Similarly, for the left-handed ($\varphi = \pi/2$) and right-handed ($\varphi = -\pi/2$) circular polarization light they are defined respectively by the following Jones vectors:
%
%
\begin{equation}\label{B7}
\tilde{\mathbf{E}}_\text{L}^c =\frac{1}{\sqrt{2}}\left(\begin{array}{c}1\\ i\end{array}\right) \,\,\text{and}\,\, \tilde{\mathbf{E}}_\text{R}^c =\frac{1}{\sqrt{2}}\left(\begin{array}{c}1\\ -i\end{array}\right).
\end{equation}
%
%
In general, for the left-handed ($\varphi = \pi/2$) and right-handed ($\varphi = -\pi/2$) elliptical polarization light in the canonical frame ($\alpha = 0$) the Jone vectors read:
%
%
\begin{equation}\label{B8}
\tilde{\mathbf{E}}_\text{L}^e =\left(\begin{array}{c}\cos\epsilon\\ i\sin\epsilon\end{array}\right) \,\,\text{and}\,\, \tilde{\mathbf{E}}_\text{R}^e =\left(\begin{array}{c}\cos\epsilon\\ -i\sin\epsilon\end{array}\right).
\end{equation}
%
%

\subsection{Equations for the polarization angle and the ellipticity}\label{appC2}
Consider a monochromatic light of the linear polarization parallel to the $Ox$ axis. This light is incident to a plane separating two material environments. The lights transmitting and reflecting at this plane will be defined by Eqs.~\eqref{B1} and~\eqref{B2}. Accordingly, the Faraday and Kerr rotation angles are determined by the value of the polarization angle $\alpha$. 

Using the complex representation for the components of the electric vector of a monochromatic light the polarization angle $\alpha$ and the ellipticity $\tan\epsilon$ are not determined by Eqs.~\eqref{B3} and~\eqref{B4}. Instead, we define the complex quantity as:
%
%
\begin{equation}
\chi = \frac{\tilde{E}_y}{\tilde{E}_x} = \frac{E_{0y}}{E_{0x}}e^{i\varphi}.
\end{equation}
%
%
Since in the canonical frame of the ellipse the Jones vector of the electric field is given by Eq.~\eqref{B8}, we should rotate it back by an angle of $\alpha$ to obtain the vector components in the global Cartesian frame $xOy$. We therefore obtain:
%
%
\begin{equation}
\left(\begin{array}{c}
\tilde{E}_x\\
\tilde{E}_y
\end{array}\right) =
\left(\begin{array}{cc}
\cos\alpha&-\sin\alpha\\
\sin\alpha & \cos\alpha
\end{array}\right)
\left(\begin{array}{c}
\cos\epsilon\\
i\eta\sin\epsilon
\end{array}\right)
\end{equation}
%
%
where $\eta=\pm 1$ is the sign for the left-handed and right-handed elliptical polarization. The quantity $\chi$ is thus determined by
%
%
\begin{equation}
\chi_\eta = \frac{\tan\alpha+i\eta\tan\epsilon}{1-i\eta\tan\alpha\tan\epsilon}.
\end{equation}
%
%
From this result it is straightforward to deduce the equations for the polarization angle and the ellipticity angle:
%
%
\begin{subequations}
\begin{align}
\tan(2\alpha_\eta) &= \frac{2\text{Re}(\chi_\eta)}{1-|\chi_\eta|^2},\label{B13}\\
\sin(2\epsilon_\eta) &=\eta\frac{2\text{Im}(\chi_\eta)}{1+|\chi_\eta|^2}.\label{B14} 
\end{align}
\end{subequations}
%
%

Now ww apply these results to determine the Kerr and Faraday rotation angles occurring at one reflection plane. In the given setup the Jones vector for the incident light is:
%
%
\begin{equation}\label{B15}
\tilde{\mathbf{E}}^{in} =\left(\begin{array}{c}\tilde{E}_x^{in}\\ 0\end{array}\right).
\end{equation}
%
%
The Jones vectors for the reflecting and transmitting lights relate to $\tilde{\mathbf{E}}^{in}$ through the reflection and transmission matrices $\mathbf{r}$ and $\mathbf{t}$, respectively, by $\tilde{\mathbf{E}}^r = \mathbf{r}\tilde{\mathbf{E}}^{in}$ and $\tilde{\mathbf{E}}^t = \mathbf{t}\tilde{\mathbf{E}}^{in}$. In particular,
%
%
\begin{subequations}
\begin{align}\label{B17}
\tilde{E}^r_x &= r_{xx}\tilde{E}_x^{in},\,\,\tilde{E}^r_y = r_{yx}\tilde{E}_x^{in},\\
\tilde{E}^t_x &= t_{xx}\tilde{E}_x^{in},\,\,\tilde{E}^t_y = t_{yx}\tilde{E}_x^{in}.
\end{align}
\end{subequations}
%
%
where $r_{\alpha\beta}$ and $t_{\alpha\beta}$ are the elements of the matrices $\mathbf{r}$ and $\mathbf{t}$. We thus obtain the expression for the $\chi$-quantity as follows:
%
%
\begin{equation}\label{B18}
\chi_\text{K} = \frac{\tilde{E}_y^r}{\tilde{E}_x^r} = \frac{r_{yx}}{r_{xx}}\,\, \text{and}\,\, \chi_\text{F} = \frac{\tilde{E}_y^t}{\tilde{E}_x^t} = \frac{t_{yx}}{t_{xx}}.
\end{equation}
%
%
By inserting $\chi_\text{K}$ and $\chi_\text{F}$ into Eq.~\eqref{B13} we obtain the Kerr and Faraday rotation angles $\theta_\text{K},\theta_\text{F}$.

\subsection{Relations between the transmission and reflection matrices to the electrical conductivity tensor}\label{appC3}
We follow the transfer matrix method to establish the expression for the transmission and reflection matrices for the system of bilayer graphene. We set up the system like the one in Ref.~\cite{Szechenyi-2016} in which the graphene layer is separating two semi-infinite mediums 1 and 2 characterized by the parameters $(\epsilon_1,\mu_1)$ and $(\epsilon_2,\mu_2)$. Ignoring the thickness of the graphene layer we can assume that the interface between the two mediums has an electrical conductivity tensor $\boldsymbol{\tilde{\sigma}} = \sigma_0\boldsymbol{\sigma}$. Therefore, the boundary conditions for the Maxwell equations at the interface read:
%
%
\begin{subequations}
\begin{align}
&\tilde{\mathbf{E}}^{in}+\tilde{\mathbf{E}}^{r} = \tilde{\mathbf{E}}^{t}\\
&\mathbf{n}\times(\tilde{\mathbf{H}}^{t}-\tilde{\mathbf{H}}^{in}-\tilde{\mathbf{H}}^{r}) = \tilde{\mathbf{J}}
\end{align}
\end{subequations}
%
%
Here $\mathbf{n}$ is the normal vector of the interface and $\tilde{\mathbf{J}}$ is the electrical current density on the interface. Because of  Ohm's law, 
%
%
\begin{equation}
\tilde{\mathbf{J}} = \boldsymbol{\tilde{\sigma}}\tilde{\mathbf{E}}^{t} = \boldsymbol{\tilde{\sigma}}(\tilde{\mathbf{E}}^{in}+\tilde{\mathbf{E}}^{r}),
\end{equation}
%
%
and the relation
%
%
\begin{equation}
\tilde{\mathbf{H}} = \sqrt{\frac{\epsilon_0\epsilon}{\mu_0\mu}}\frac{\mathbf{k}\times\tilde{\mathbf{E}}}{k}
\end{equation}
%
%
we identify the expression for the transmission matrix:~\cite{Leyva-2016}
\begin{equation}
\mathbf{t} = 2\sqrt{\frac{\epsilon_0\epsilon_1}{\mu_0\mu_1}}\left[\left(\sqrt{\frac{\epsilon_0\epsilon_1}{\mu_0\mu_1}}+\sqrt{\frac{\epsilon_0\epsilon_1}{\mu_0\mu_1}}\right)\mathbf{I}+\boldsymbol{\tilde{\sigma}}\right]^{-1}.
\end{equation}
Here $\mathbf{I}$ is the identity matrix. The reflection matrix $\mathbf{r}$ is determined via the relation $\mathbf{r}+\mathbf{t} = \mathbf{I}$.

Assuming $\mu_1=\mu_2 = 1$ and noting the definition of the refractive index $n_{1,2} \approx \sqrt{\epsilon_{1,2}}$, we have:
%
%
\begin{equation}
\mathbf{t} = 2\left[\left(1+n_{21}\right)\mathbf{I}+\bar{\boldsymbol{\sigma}}\right]^{-1}.
\end{equation}
%
%
where $n_{21} = n_2/n_1$, $\bar{\boldsymbol{\sigma}} = (4\pi\hbar\alpha/n_1e^2)\boldsymbol{\tilde{\sigma}} = (\pi\alpha/n_1)\boldsymbol{\sigma}$, and  $\alpha = e^2/(4\pi\hbar\epsilon_0c)\approx 1/137$ is the fine-structure constant. Proceeding with further calculations, we obtain:
%
%
\begin{equation}\label{B25}
\mathbf{t} = \frac{2}{\Delta}\begin{pmatrix}
1+n_{12}+\bar{\sigma}_{yy} & -\bar{\sigma}_{xy}\\
-\bar{\sigma}_{yx} & 1+n_{12}+\bar{\sigma}_{xx}
\end{pmatrix}
\end{equation}
%
%
and
%
%
\begin{equation}\label{B26}
\mathbf{r} = \frac{2}{\Delta}\begin{pmatrix}
1+n_{12}+\bar{\sigma}_{yy}-\frac{\Delta}{2} & -\bar{\sigma}_{xy}\\
-\bar{\sigma}_{yx} & 1+n_{12}+\bar{\sigma}_{xx}-\frac{\Delta}{2}
\end{pmatrix}
\end{equation}
%
%
where $\Delta = [(1+n_{21})+\bar{\sigma}_{xx}][(1+n_{21})+\bar{\sigma}_{yy}]-\bar{\sigma}_{xy}\bar{\sigma}_{yx}$.

\subsection{Circular dichroism}
The circular dichroism is a quantity used to measure the dependence of the light absorption on the left-handed and right-handed polarization:
%
%
\begin{equation}
\text{CD} = \frac{A_\text{L}-A_\text{R}}{A_\text{L}+A_\text{R}},
\end{equation}
%
%
where $A_\text{L/R}$ are the absorptances for left and right polarized light. These quantities are determined through the reflectance $R$ and transmittance $T$ by $A_\text{L/R} = 1-(R_\text{L/R}+T_\text{L/R})$. Here the reflectance and transmittance are determined by 
%
%
\begin{subequations}
\begin{align}
R &= \frac{(\mathbf{r}\tilde{\mathbf{E}}^{in})^\dagger(\mathbf{r}\tilde{\mathbf{E}}^{in})}{(\tilde{\mathbf{E}}^{in})^\dagger\tilde{\mathbf{E}}^{in}},\\
T &= \frac{(\mathbf{t}\tilde{\mathbf{E}}^{in})^\dagger(\mathbf{t}\tilde{\mathbf{E}}^{in})}{(\tilde{\mathbf{E}}^{in})^\dagger\tilde{\mathbf{E}}^{in}}.
\end{align}
\end{subequations}
%
%
In particular, with the notice of the Jones vector given in Eq.~\eqref{B7} the absorptances of the left/right-handed circular light are given by:
%
%
\begin{subequations}
\begin{align}
A_\text{L} &= \text{Re}\left[t_{xx}+i(t_{xy}-t_{yx})+t_{yy}\right],\\
A_\text{R} &= \text{Re}\left[t_{xx}-i(t_{xy}-t_{yx})+t_{yy}\right].
\end{align}
\end{subequations}
%
%
The formula for the CD therefore reads:
%
%
\begin{equation}\label{B32}
\text{CD} = \frac{\text{Im}(\bar{\sigma}_{xy}-\bar{\sigma}_{yx})}{\text{Re}[2(1+n_{21})+\bar{\sigma}_{xx}+\bar{\sigma}_{yy}]}.
\end{equation}
%
%

For mediums 1 and 2 being vacuum, we can set $n_1=n_2 =n_{21} =1$.

\bibliography{bibliography}
\end{document}